\documentclass[twocolumn,nopacs,prl,amsfonts,amsmath,amssymb,floatfix]{revtex4} 
\usepackage{color}
\usepackage{hhline}	
\usepackage{mathrsfs}
\usepackage{graphicx}
\usepackage{dcolumn}
\usepackage{bm}% bold math
\usepackage{multirow}
\usepackage{booktabs}
\usepackage{afterpage}
\usepackage{amsmath}
\usepackage{braket}
\usepackage{here}

\begin{document}

\title{Possible correlated insulating states in magic-angle twisted bilayer graphene under strongly competing interactions}

\author{Masayuki Ochi}
\author{Mikito Koshino}
\author{Kazuhiko Kuroki}
\affiliation{Department of Physics, Osaka University, Machikaneyama-cho, Toyonaka, Osaka 560-0043, Japan}

\date{\today}
\begin{abstract}
We investigate correlated insulating states in magic-angle twisted bilayer graphene (TBG) by the exact diagonalization method applied to the extended Hubbard model with interaction parameters recently evaluated in the realistic effective model. Our model can handle the competing interactions among Wannier orbitals owing to their significant overlap, which is a crucial but overlooked aspect of the magic-angle TBG.
We propose two candidates for the correlated insulating states: spin- and valley-ferromagnetic band insulator and the Dirac semimetallic state for two flavors with peculiar renormalization, where a flavor denotes a combined degree of freedom with spin and valley. One of the important consequences for the latter candidate is that it allows van Hove singularity near half-filling of the whole band structure (i.e. near the Dirac points) to play some role in superconductivity.
The consistency between the two flavor degrees of freedom for the Dirac semimetallic state and the two-fold degeneracy of the Landau level observed in the experiment is also noteworthy.
\end{abstract}
%\pacs{}

\maketitle

\begin{figure}
\begin{center}
\includegraphics[width=7.5 cm]{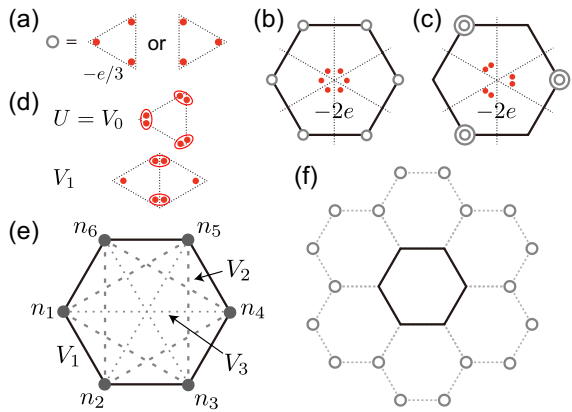}
\caption{(a) Wannier function with a three-peaked structure. (b)(c) Two (nearly) degenerate electron configurations where single and double circles denote one or two electrons on each vertex, respectively. (d) Direct Coulomb interaction between the Wannier orbitals. (e) Single hexagon with the definition of $V_i$ ($i=1,2,3$) and the electron number $n_i$ ($i=1,\dots,6$) at each vertex. (f) Single hexagon embedded in the uniform background.}
\label{fig:1}
\end{center}
\end{figure}

A recent study revealed the emergence of unconventional superconductivity in magic-angle twisted bilayer graphene (TBG)~\cite{tbg_sc}. Because of its unique electronic structure where the band structure with a very narrow width hosts the superconductivity near the correlated insulating states at some electron fillings~\cite{tbg_ins},
a lot of studies have been carried out to investigate its non-trivial electronic structure.
However, the nature of the superconducting phase~\cite{Guo_sc,Liu_RPA,Huang,Baskaran,Kennes,Xu_sc,Fidrysiak_sc,Roy,Peltonen_sc,WuEP,Ray_sc,Dodaro,Isobe,icvw}
and the correlated insulating state~\cite{Huang,Pizarro,Irkhin,Roy,Padhi_Wigner,Xu_kekule,Liu_RPA,Dodaro,Po,Isobe,weakMI,icvw} are under active debate.
Many possibilities of the exotic electronic structures reflect the fact that TBG can be a playground for various kinds of rich physics, but it partially comes from a difficulty in identifying the effective low-energy model used for dealing with strong correlation effects in TBG.
This is because the moir{\'e} pattern in TBG with small twist angles requires a very large system size, which makes a usual first-principles electronic structure calculation very difficult.
Very recently, Koshino {\it et al.} constructed a realistic effective model by taking into account hybridization of the Bloch states in the original graphene and the corrugation effect of the stacked layers~\cite{model_Koshino}.
They evaluated not only the hopping amplitudes for the tight-binding model but also the interaction parameters.
Their study revealed intriguing features of the electronic structure in magic-angle TBG.
For example, the inter-site interaction including the exchange interaction has a comparable magnitude to that of the on-site interaction due to a significant overlap among the Wannier orbitals.
This is due to a three-peaked form of the Wannier orbitals~\cite{Po,model_Kang,model_Koshino} where these peaks are overlapped with those of other Wannier orbitals as we shall see in Figs.~\ref{fig:1}(a)--\ref{fig:1}(d). As a result, many-body states at the band filling for the correlated insulating states (i.e., quarter-filling for the whole band structure) suffer from unusual degeneracy as mentioned later.
This aspect is of crucial relevance to the mysterious nature of the correlated insulating state and the superconducting phases close to it in magic-angle TBG, while this aspect has not been considered appropriately in the previous theoretical studies.

In this Rapid Communication, we investigate the nature of the correlated insulating states in magic-angle TBG by the exact diagonalization method applied to the extended Hubbard model with interaction parameters evaluated in the realistic effective model~\cite{model_Koshino}.
Here, the characteristic interaction owing to the three-peaked structure of the Wannier orbitals is a robust feature of small-angle TBGs~\cite{Po,model_Kang,model_Koshino} while the band structure somewhat varies depending on the models~\cite{WuEP,model_Koshino}.
Based on our analysis that appropriately takes into account competing interactions as a fundamental aspect of TBG, we propose two candidates for the correlated insulating states: spin- and valley-ferromagnetic band insulator and the Dirac semimetallic state for two flavors with peculiar renormalization, where a flavor denotes a combined degree of freedom with spin and valley. A significant consequence for the latter possibility is that it allows van Hove singularity (vHs) to play some role in superconductivity even when vHs exists near the Dirac points (i.e., half-filling of the whole band structure) as in the model derived in Ref.~\cite{model_Koshino}.
The consistency between the two flavor degrees of freedom for the Dirac semimetallic state and the two fold degeneracy of the Landau level observed in the experiment~\cite{tbg_sc} is also noteworthy as described later.

First, we show a step-by-step simplification of the effective model derived in Ref.~\cite{model_Koshino}.
The Wannier orbitals of TBG have the three-peaked structure~\cite{Po,model_Kang,model_Koshino} as presented in Fig.~\ref{fig:1}(a), where the Wannier orbital is denoted by an open circle while a red small solid circle with $-e/3$ charge presents a single peak. As shown in Ref.~\cite{model_Koshino}, Wannier orbitals constitute a hexagonal lattice.
For example, in Figs.~\ref{fig:1}(b) and \ref{fig:1}(c), each vertex on a hexagon corresponds to the Wannier orbital, where single and double circles in the figure represent one or two electrons on each site, respectively.
Since a single Wannier orbital can be regarded as a set of three $-e/3$ charges forming a triangle, the centers of the hexagons both in Figs.~\ref{fig:1}(b) and \ref{fig:1}(c) have $-e/3\times 6=-2e$ charge.
One of the important findings in Ref.~\cite{model_Koshino} is that the intra- and inter-site direct Coulomb repulsion is very well approximated by the simple point-charge model for the $-e/3$ charge~\cite{note_pointcharge}.
Because one cannot distinguish two electron configurations presented in Figs.~\ref{fig:1}(b) and \ref{fig:1}(c) in terms of the distribution of $-e/3$ charges, these two states should have the (almost) same energy with respect to the intra- and inter-site direct Coulomb repulsion.

This degeneracy is one of the most important aspects of the electronic structure in magic-angle TBG. 
For example, it is not true that the Mott-like uniform distribution of electrons at each site presented in Fig.~\ref{fig:1}(b) always has a much lower energy than other states, which is expected by only considering the on-site repulsion at each site.
To keep this degeneracy under the restriction on the reachable system size for exact diagonalization, we limited the direct Coulomb repulsion among $-e/3$ fractional charges to the `on-site' one, i.e., the fractional charges feel the `on-site' interaction at the center of the hexagon.
 This assumption corresponds to $V_0(=U):V_1:V_2:V_3=3:2:1:1$ without more distant interaction as explained in Ref.~\cite{model_Koshino} [cf. Fig.~\ref{fig:1}(d) where the definition of $V_i$ ($i=1,2,3$) is shown in Fig.~\ref{fig:1}(e)].
Under this approximation, the energy per hexagon shown in Fig.~\ref{fig:1}(e) is determined by a sum of the electron numbers at the six vertices: $\sum_{i=1}^6 n_i$~\cite{footnote_proof1}, which guarantees the degeneracy of two states shown in Figs.~\ref{fig:1}(b) and \ref{fig:1}(c).
This approximation is a minimum one to deal with the complex degeneracy inherent to this system, and we note that a naive truncation of the interaction terms at some distance does not keep this degeneracy.

The model solved in this Rapid Communication is presented in Fig.~\ref{fig:1}(f), where a single hexagon is surrounded by six hexagons with a fixed occupation: one electron per site, which corresponds to the uniform distribution for quarter-filling. Here, we took the nearest-neighbor hopping $t$ inside the central hexagon but neglected the other hopping terms inside the hexagon since they are sufficiently smaller than $t$~\cite{footnote_hopping}. We also took the inter-nearest-neighbor-site exchange interaction $J_1$, which was evaluated to be $\sim -U/5$~\cite{model_Koshino}.
This model has two spin and two valley degrees of freedom, and we shall call these degrees of freedom {\it flavor} (i.e., this model has four flavors). Here, a different valley corresponds to a time-reversal counterpart of the Wannier functions.
We note that the hopping and exchange terms act only between the same flavor. On-site Hund's coupling between different valleys is negligible~\cite{model_Koshino}. Our effective Hamiltonian reads
\begin{align}
\mathcal{H} = t \sum_{i=1}^6 \sum_{\alpha=1}^4 [ \hat{c}^{\dag}_{i+1,\alpha} \hat{c}_{i,\alpha} + \mathrm{H.c.} ] + \frac{V_0}{2} \sum_{I} \sum_{\alpha \neq \beta} \hat{n}_{I,\alpha} \hat{n}_{I,\beta} \nonumber \\
+ \frac{V_1}{2} \sum_{I_1,I_2}^{\mathrm{Nearest}} \sum_{\alpha, \beta} \hat{n}_{I_1,\alpha} \hat{n}_{I_2,\beta}
+ \frac{V_2}{2} \sum_{I_1,I_2}^{\mathrm{NN}} \sum_{\alpha, \beta} \hat{n}_{I_1,\alpha} \hat{n}_{I_2,\beta} \nonumber \\
+ \frac{V_3}{2} \sum_{I_1,I_2}^{\mathrm{NNN}} \sum_{\alpha, \beta} \hat{n}_{I_1,\alpha} \hat{n}_{I_2,\beta} 
+ \frac{J_1}{2} \sum_{i_1,i_2}^{\mathrm{Nearest}} \sum_{\alpha} \hat{n}_{i_1,\alpha} \hat{n}_{i_2,\alpha},
\end{align}
with $V_0=U$, $V_1=2U/3$, $V_2=V_3=U/3$, and $J_1=-U/5$~\cite{footnote_int_flav}.
The index $i$ runs over six lattice sites of the central hexagon, and $I$ runs over all the sites including the outer hexagons.
$\alpha$ and $\beta$ are flavor indices, NN and NNN mean next-nearest and next-next-nearest sites, respectively, and $\hat{c}^{\dag}_7=\hat{c}^{\dag}_1$.
As we shall see later, the six `background' hexagons with the contribution of the direct Coulomb repulsion are necessary to understand a fundamental excitation of magic-angle TBG in an intuitive manner.
In order to consider the competition between $t$ and $J_1$ as discussed later, we took the $J_1$ terms only inside the central hexagon in the same manner as $t$.
We considered the quarter-filled case: six electrons on six sites in the central hexagon.
Since a dimension of the block (described later) of our Hamiltonian is up to 8,100, we performed full diagonalization.

\begin{figure}
\begin{center}
\includegraphics[width=8.2 cm]{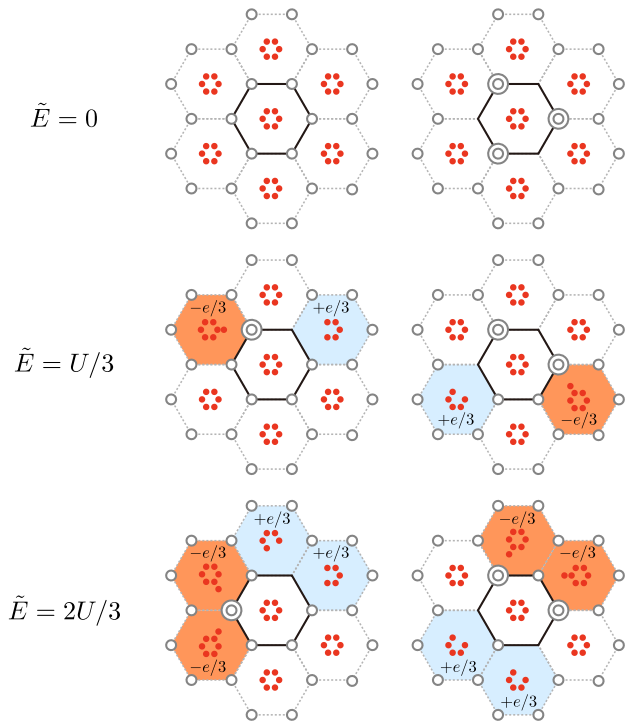}
\caption{Three lowest energies obtained using $t=J_1=0$ with representative configurations. Red small filled circles denote $-e/3$ fractional charge while single and double circles denote one ($-e$) or two ($-2e$) electrons on each vertex, respectively (cf. Fig.~\ref{fig:1}).}
\label{fig:2}
\end{center}
\end{figure}

Now we move on to the results obtained by our analysis.
Figure~\ref{fig:2} presents representative states of the three lowest energies, obtained by taking $t=J_1=0$ as a first step.
$\tilde{E}$ denotes the total energy subtracted with a constant to have $\tilde{E}=0$ for the lowest energy here.
Red small filled circles in Fig.~\ref{fig:2} denote $-e/3$ fractional charge while single and double circles denote one ($-e$) or two ($-2e$) electrons on each vertex, respectively (cf. Fig.~\ref{fig:1}). While all the lowest energy states with $\tilde{E}=0$ have six fractional charge at the center of each hexagon, the electron configurations of the excited states can be regarded as that with additional or reduced fractional charge at some hexagons as denoted with $-e/3$ or $+e/3$ in Fig.~\ref{fig:2}, respectively. As a result, electron configurations for $\tilde{E}=U/3$ represent a kind of elementary excitation with a fractional charge~\cite{model_Koshino,note_frac}. In fact, electron configurations for $\tilde{E}=2U/3$ can be regarded as the states with two elementary excitations. A pair of two charges with different signs reminds us of the exciton or the doublon-holon pair.

\begin{figure}
\begin{center}
\includegraphics[width=8.2 cm]{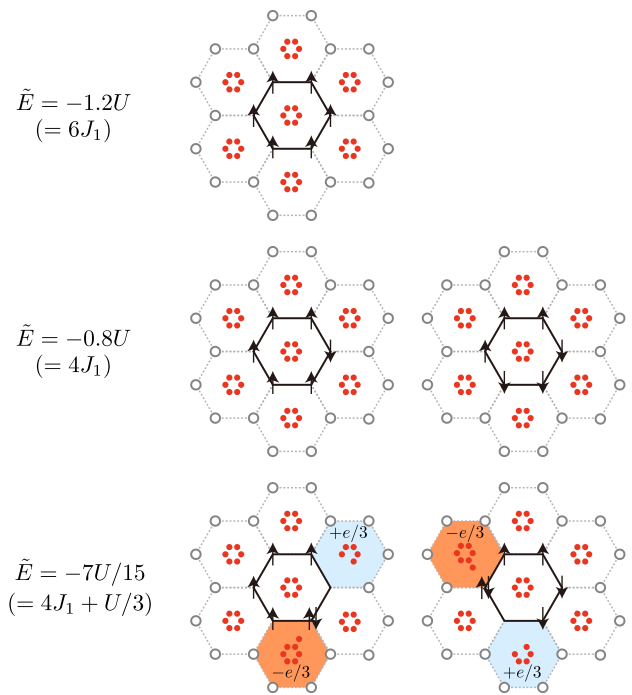}
\caption{Three lowest energies obtained using $t=0$ with representative configurations. Arrows with the same direction denote electrons with the same flavor. The meaning of other symbols is the same as Fig.~\ref{fig:2}.}
\label{fig:3}
\end{center}
\end{figure}

By including $J_1=-U/5$, we obtained electron configurations shown in Fig.~\ref{fig:3}.
The most stable states consist of the electrons with a single flavor, where six $J_1$ bonds are active.
The second-lowest states have some flipped flavors, but four $J_1$ bonds remain to be active.
The lowest and the second-lowest states have the most favorable electron configurations for the $V_i$ terms: one electron per site as we have seen in Fig.~\ref{fig:2}. The third-lowest states show that the elementary excitation mentioned above remains a good unit to understand the energy spectrum of this system~\cite{note_fig3}.

\begin{figure}
\begin{center}
\includegraphics[width=8.5 cm]{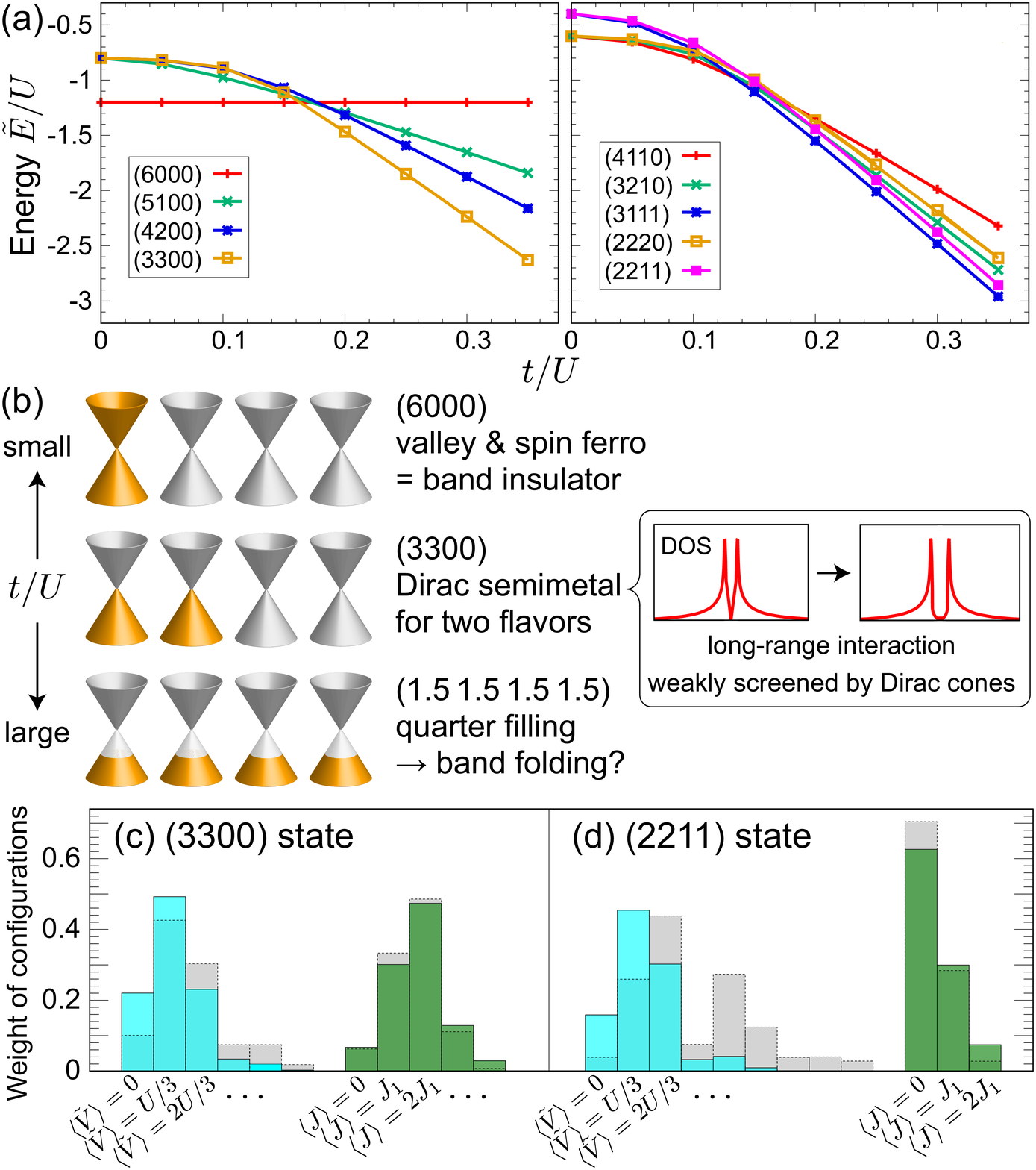}
\caption{(a) Lowest energies obtained for each block of Hamiltonian as a function of $t/U$. A set of four integers ($n_{\alpha_1}$ $n_{\alpha_2}$ $n_{\alpha_3}$ $n_{\alpha_4}$) denotes a single block consisting of the states with the electron numbers $n_{\alpha_i}$ for each flavor $\alpha_i$. (b) Schematic picture for three candidates of correlated insulating states in magic-angle TBG. For the Dirac semimetallic state, a schematic picture for the renormalized DOS is also shown. (c)(d) Histogram of the weight of electron configurations for the lowest eigenstates in the (3300) and (2211) blocks for each panel using $t=U/5$. }
\label{fig:4}
\end{center}
\end{figure}

Finally, we took all the terms in our Hamiltonian.
Figure~\ref{fig:4}(a) presents the lowest energies obtained for each block of Hamiltonian as a function of $t/U$. Here we define an index of a block of our Hamiltonian, ($n_{\alpha_1}$ $n_{\alpha_2}$ $n_{\alpha_3}$ $n_{\alpha_4}$), using the electron number $n_{\alpha_i}$ for each flavor $\alpha_i$~\cite{note_block}.
In the left panel of Fig.~\ref{fig:4}(a), there is no energy gain by the hopping terms for the (6000) state because electrons are fully occupied in a single flavor and so cannot move by the Pauli principle (see the lowest eigenstate shown in Fig.~\ref{fig:3}). 
Therefore, while the (6000) state is the most stable for $t/U=0$ among all the blocks of Hamiltonian, other states become more stable by increasing $t/U$.
For a sufficiently large $t/U$, an equal distribution for four flavors is expected to be the most stable to gain the largest kinetic energy profit, while our small model cannot represent such an equal distribution, (1.5 1.5 1.5 1.5).
In this sense, competition among some blocks near this equal distribution, such as (3111), (3210), and (2211), for $t/U > 0.3$ can be regarded as a finite-size effect~\cite{footnote_size}.
For such a quarter-filled state for four flavors, (1.5 1.5 1.5 1.5), the system has a chance to become insulating if the charge-density order appears and the Brillouin zone is folded with opening a gap~\cite{icvw}, but a detailed analysis on it is beyond the range of our study.
For intermediate $t/U$, it is rather non-trivial which state is the most stable under the strong competition.
To explore the possible insulating states observed in experiment, it is worth noting that the (3300) state, one of the competitors, has a possibility to become an insulator-like state for the following reason.
The (3300) occupation in our model corresponds to the electronic structure where two flavors are half-filled.
Because $t/U\sim 0.2$ where the non-trivial competition was observed in our model is expected to yield a relatively strong correlation, the corresponding occupation still has a chance to be the ground state in the real system nevertheless a detailed band structure is not reproduced in our small model.
In the real TBG, the Fermi level resides at the Dirac points for half-filling where the density of states (DOS) vanishes. While it is still semimetallic, it is known that the presence of the long-range Coulomb interaction owing to the weak screening effect by Dirac points can anomalously increase the Fermi velocity of the Dirac cones, and DOS near the Dirac points can be suppressed~\cite{graphene_renorm1,graphene_renorm2,graphene_renorm3,graphene_renorm4,graphene_renorm5}.
In addition to this peculiar renormalization of Dirac cones, a very sharp vHs close to the Dirac points in magic-angle TBG~\cite{model_Koshino} can make the system look like an insulator.
For the Dirac semimetallic state for two flavors, there is another interesting consistency with the experiments. Namely, the quantum oscillation experiment reported that the Landau level degeneracy near the quarter-filling is two-fold~\cite{tbg_sc} rather than four-fold as naturally expected from the four flavor degrees of freedom in this system~\cite{footnote_symbre}.
This is clearly consistent with the Dirac semimetallic state for two flavors shown in Fig.~\ref{fig:4}(b).

These three possibilities for correlated insulating states in the small, intermediate, and large $t/U$ regimes are summarized in Fig.~\ref{fig:4}(b).
Also we do not exclude the possibility of the phase separation of the flavor-ferromagnetic regions with different flavors since the stability of the (3300) state can lead to such a phase separation.
We will leave further analysis on the competing nature of the ground state in larger systems for the future since it requires high computational cost. If one wants to treat a larger size while keeping the $C_3$ and AB sublattice symmetries, a system with at least 12 sites is required.

While detailed analysis is difficult owing to the finite-size effect, we mention a few more observations of our model. Figures~\ref{fig:4}(c) and \ref{fig:4}(d) present a histogram of the weight of electron configurations for the lowest eigenstates in the (3300) and (2211) blocks, respectively, using $t=U/5$. $\langle \tilde{V} \rangle$ and $\langle J \rangle$ denote the expectation values of the $V_i$ ($i=0,1,2,3$) and $J_1$ terms, respectively, where the constant is subtracted for $\langle \tilde{V} \rangle$ in the same manner as $\tilde{E}$. The main difference between these two states is that the (3300) state actually obtains an energy gain by $J_1$ while it is much weaker in the (2211) state. 
The gain from $J_1$ also leads to the stabilization of the uniformly distributed states as shown in the top left in Fig.~\ref{fig:2} than the localized states as shown in the top right in Fig.~\ref{fig:2} since the former has a chance to be stabilized by $J_1$. As a matter of fact, the (3300) eigenstate with $t=U/5$ has $\langle \bar{n}_A \bar{n}_B \rangle = 0.82$, where $\bar{n}_A=(n_1+n_3+n_5)/3$ and $\bar{n}_B=(n_2+n_4+n_6)/3$, which is closer to the uniform distribution ($\langle \bar{n}_A \bar{n}_B \rangle=1$) than the localized one ($\langle \bar{n}_A \bar{n}_B \rangle=0$). The competition between the exchange interaction and the stabilization by hybridization through hopping terms might be a key to understand the nature of the correlated states. We stress once again the reason why these two factors compete: the exchange interaction favors the flavor-ferromagnetic band insulator while electrons cannot move there.

As for the superconductivity, an important consequence of our study is that vHs can play some role in stabilizing superconductivity even when it resides near the Dirac points.
While some studies proposed that vHs near quarter-filling plays a key role in superconductivity (e.g. ~\cite{Liu_RPA,Isobe,Kennes}), some realistic calculation results in vHs very close to the Dirac points~\cite{model_Koshino}.
Our proposal for the correlated state with the half-filled band dispersions for two flavors enables the Fermi energy to reside near vHs as shown in Fig.~\ref{fig:4}(b).
Therefore, some scenarios for superconductivity making use of vHs is justified even when vHs exists near the Dirac points.
Here, vHs can be involved with superconductivity not only from the viewpoint of the electron correlation, but also through enhancement of the electron-phonon coupling~\cite{Peltonen_sc,Isobe,WuEP} by its large DOS. Because the lattice relaxation plays a crucial role for TBG such as making separated band dispersion through gap opening~\cite{gap1,model_Nam}, the phonon can have a sizable effect also on the correlated states.

To summarize, we performed exact diagonalization of the extended Hubbard model for magic-angle TBG and proposed two new candidates for the correlated insulating states: spin- and valley-ferromagnetic band insulator and the Dirac semimetallic state for two flavors with peculiar renormalization. For the latter possibility, vHs near the Dirac points can be involved with superconductivity. The consistency between the two flavor degrees of freedom for the Dirac semimetallic state and the two-fold degeneracy of the Landau level observed in the experiment is also remarkable.
Our study will be an important basis for studying unusual correlation effects including the superconductivity in magic-angle TBG.

\begin{acknowledgments}
MO acknowledges the financial support of JSPS KAKENHI Grant No. JP18K13470. MK is supported by JSPS KAKENHI Grant No. JP17K05496.
KK is supported by JSPS KAKENHI Grant No. JP18H01860.
\end{acknowledgments}


\begin{thebibliography}{999}
\bibitem{note_pointcharge} Direct Coulomb interaction between two point charges is assumed to be proportional to $1/r$ where $r$ is the distance between the point charges. When two point charges exist in the same place, the direct Coulomb energy between them is evaluated as a finite constant. For detail, see Ref.~\cite{model_Koshino}.
\bibitem{footnote_proof1} A sum of the $V_0,V_1,V_2,V_3$ contributions per hexagon is $V_0(N^2-N)/6$ where $N=\sum_{i=1}^6 n_i$ provided that $V_0:V_1:V_2:V_3=3:2:1:1$. For example, the $V_0$ contribution per hexagon equals $(V_0/3)\sum_{i=1}^6 n_i(n_i-1)/2$, where the factor $1/3$ comes from the fact that each vertex is shared with three neighboring hexagons. The expression $V_0(N^2-N)/6$ corresponds to a product of the number of the pairs of fractional charge $-e/3$ in this hexagon and the on-site interaction per fractional-charge pair, $V_0/3$ [see Fig.~\ref{fig:1}(d)].
\bibitem{footnote_hopping} In Ref.~\cite{model_Koshino}, the transfer integrals inside the central hexagon except $t$ are at least nine times smaller than $t$ (e.g. $t_4$ in Ref.~\cite{model_Koshino}). Although there are distant transfer integrals with a larger amplitude, which are around three or four times smaller than $t$ (such as $t_2$ in Ref.~\cite{model_Koshino}), they cannot be considered in our model within the reachable small system size for exact diagonalization.
\bibitem{footnote_int_flav} The interaction parameters $V_i$ and $J$ depend on neither spin nor valley. This is because the Wannier functions with different spins are completely the same and those with different valleys are conjugate to each other since they are time-reversal counterparts (see Ref.~\cite{model_Koshino} for more details).
\bibitem{note_frac} Because $-e/3$ and $+e/3$ hexagons are not neighboring for $\tilde{E}=U/3$ states as shown in Fig.~\ref{fig:2}, the energy for this elementary excitation does not change even when one includes the nearest-neighbor direct Coulomb interaction between $-e/3$ charges, while two elementary excitations existing in $\tilde{E}=2U/3$ states can weakly interact in that case.
\bibitem{note_fig3} We note that the expectation value of the $V_i$ terms is solely determined by the number of the fractional charge at the center of each hexagon: $\sum_{i=1}^6 n_i$ in Fig.~\ref{fig:1}(e) (cf.~\cite{footnote_proof1}).
\bibitem{note_block} We note that, because the (6000) block and the (0600) block etc.~are equivalent owing to the equivalency of the flavors, we do not specify which spin and valley each $\alpha_i$ corresponds to.
\bibitem{footnote_size} The one-electron energies for the kinetic energy terms without the interaction terms for the hexagonal ring are $-2t$, $-t$ with two-fold degeneracy, $t$ with two-fold degeneracy, and $2t$. Therefore, for example, both the (31) and (22) filling for two flavors yield the total kinetic energy $-4t$ when neglecting the interaction terms. Because of a stronger stabilization in (31) configuration through the exchange term, apparently equal distribution (22) cannot be more stable than (31) distribution even with a very large $t/U$ as an artifact of our six-site model.
\bibitem{footnote_symbre} A necessity of the symmetry breaking with respect to the flavor degrees of freedom was pointed out through the symmetry analysis presented in Ref.~\cite{Po}.
\end{thebibliography}
\end{document}